\documentclass{article}
\usepackage{spconf,amsmath,graphicx}
\usepackage{multirow}
\usepackage{enumitem}
\setlist{nosep, leftmargin=14pt}

\usepackage{mwe} 
\usepackage{booktabs}
\usepackage{multirow}
\usepackage[colorlinks=true, allcolors=green]{hyperref}
\usepackage{amsmath, amssymb}

\title{UMedNeRF: Uncertainty-aware Single View Volumetric Rendering for Medical Neural Radiance Fields}
\name{
Jing Hu$^{1}$,
Qinrui Fan$^{1}$,
Shu Hu$^{2}$,
Siwei Lyu$^{3}$, 
Xi Wu$^{1,\dagger}$,
Xin Wang$^{4,\dagger}$ \thanks{ $\dagger$ Corresponding authors (wuxi@cuit.edu.cn, xwang56@albany.edu).}
}
\address{
 $^{1}$ Chengdu University of Information Technology, Chengdu, China\\
 $^{2}$ Purdue University, IN, USA \\
  $^{3}$ University at Buffalo, State University of New York, NY, USA \\
 $^{4}$ University at Albany, State University of New York, NY, USA
}
\begin{document}
%
\maketitle
\begin{abstract}
In the field of clinical medicine, computed tomography (CT) is an effective medical imaging modality for the diagnosis of various pathologies. Compared with X-ray images, CT images can provide more information, including multi-planar slices and three-dimensional structures for clinical diagnosis. However, CT imaging requires patients to be exposed to large doses of ionizing radiation for a long time, which may cause irreversible physical harm. In this paper, we propose an Uncertainty-aware MedNeRF (UMedNeRF) network based on generated radiation fields. This network can learn a continuous representation of CT projections from 2D X-ray images by obtaining the internal structure and depth information and using multi-task adaptive loss weights to ensure the quality of the generated images. Our model is trained on publicly available knee and chest datasets, and we show the results of CT projection rendering with a single X-ray and compare our method with other methods based on generated radiation fields.
\end{abstract}
\section{Introduction}
\label{sec:intro}

Computed Tomography (CT) plays a vital role in medical imaging as it provides detailed cross-sectional images that are essential for accurate diagnosis and treatment planning. However, CT scans involve higher levels of radiation exposure for patients compared to X-rays. By utilizing X-ray data for CT reconstruction, it is possible to reduce the radiation dose received by patients\cite{lo2012extraction}.

Deep learning models \cite{wang2024artificial} such as 3D convolutional neural networks, generative adversarial networks (GAN), and variational autoencoders can be applied to the reconstruction of medical images like CT and MRI \cite{maken20232d}. However, these methods often demand high-performance computing resources due to the computational complexity in three-dimensional space. 
Furthermore, they typically require extensive training data and achieving satisfactory results can be challenging \cite{wang2023deep}. Neural radiance fields (NeRF) \cite{mildenhall2021nerf} method has yielded surprising results in traditional 3D image reconstruction, but it cannot be directly applied to medical image reconstruction due to the unique internal details of medical images \cite{wang2024neural}.


\begin{figure}[t] %
\centering
\includegraphics[width=0.48\textwidth]{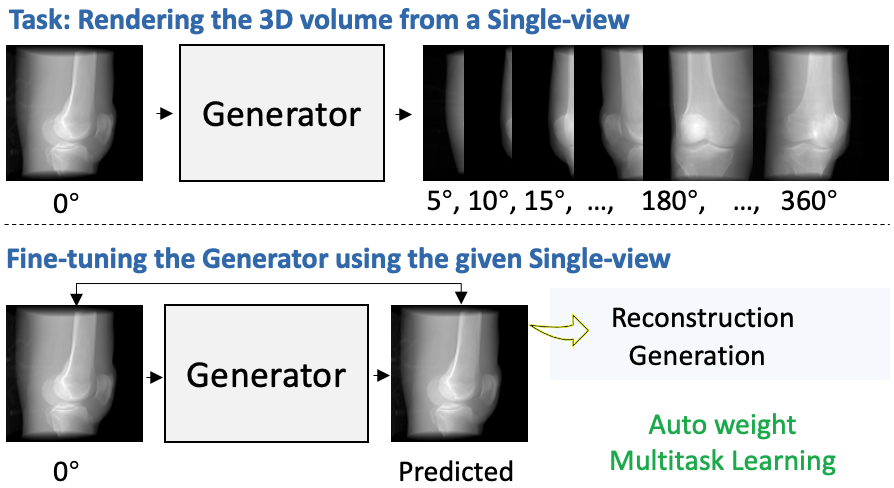}
\vspace{-5mm}
\caption{\small Overview of the task of UMedNeRF. \textbf{Top:} The goal is to render the 3D volume from a single view. \textbf{Bottom:} Given the trained Generator, it is further fine-tuned for the final rendering process via a multitask learning method.}
\label{figintro}
\end{figure}

\begin{figure*}[t] %
\centering
\includegraphics[width=1\textwidth]{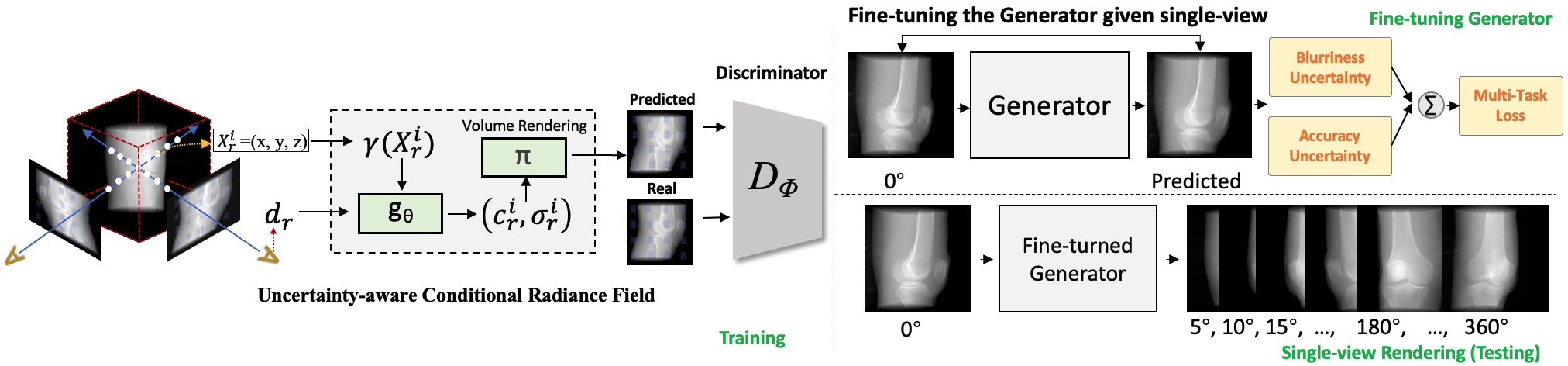}
\vspace{-8mm}
\caption{\small Overview of our Uncertainty-aware MedNeRF. \textbf{Left:} Network structures. \textbf{Right:} (Top) Fine-tuning the generator for balancing the Blurrinenss and Accuracy. (Bottom) Rendering the whole 3D volume (each view taken at a 5-degree interval) using a single view slide.}
\label{figov}
\vspace{-3mm}
\end{figure*}

MedNeRF \cite{corona2022mednerf}, a GAN-based Raditonal Filed, successfully untangles the surface shape, volumetric depth, and internal anatomical structures from 2D images by learning a function that assigns a radiance value to each pixel. However, due to the nature of GAN network architecture for image generation and using a single X-ray for CT projection reconstruction during image generation, the network may overfit to the single X-ray, leading to an imbalance between image clarity and accuracy. Thus, during the single-view volumetric rendering stage, MedNeRF fine-tuned the generator using the input view by jointly optimizing the reconstruction and the generator. Two losses were leveraged in MedNeRF:
the reconstruction loss is based on the disparity between the predicted and actual CT projections, while the generative loss is calculated using the variance between the predicted and actual radiance fields. 
The hyperparameters for balancing two losses are typically configured manually, 
which can pose challenges in achieving the optimal trade-off between image blurriness and accuracy. 

To address this challenge, in this paper, we propose Uncertainty-aware MedNeRF (\textbf{UMedNeRF}) to automatically learn hyperparameters for clarity and precision balancing problems (See Fig. \ref{figintro}). 
We propose to use an uncertainty-aware conditional radiance field that simultaneously weighs multiple loss functions automatically in the multi-task settings and generates the predicted patch for single-view volumetric rendering.
The experiments illustrate that our method significantly improves the model's reconstruction abilities while accelerating the convergence rate during testing.

\section{METHODOLOGY}



\noindent\textbf{Overview.} The overview of the framework of our UMedNeRF is shown in Fig. \ref{figov} (Left). We use the same network structures in the MedNeRF \cite{corona2022mednerf}. The whole structure consists of a radiance field network and a multi-scale patch-based discriminator. 
The 2D image is encoded as 3D position $X= (x,y,z)$, and the viewing angle direction information $d$ is fed into generator $g$ as input. Network outputs volume density $\sigma$ and pixel value $c$. At the same time, in order to learn the high-frequency characteristics of the input information, the input position information is mapped to a two-dimensional representation $\gamma(X)$, as Eq. (\ref{2dimensinal}). Finally, the output $(c, \sigma)$ is rendered as a predicted patch by Volume Rendering. The multi-scale patch-based discriminator takes in a pair of images and outputs a scalar value indicating whether the pair is real or generated.
\begin{equation}
\gamma(X)=(\ldots, \sin \left(2^i \pi X\right), \cos \left(2^i \pi X\right), \ldots)
\label{2dimensinal}
\end{equation}





\noindent\textbf{Uncertainty-aware Generative Radiance Fields.} Compared with natural images, medical images such as CT scans and MRI bring unique challenges to 3D reconstruction due to the problems of unclear boundaries and complex internal structures. It is difficult to establish a continuous representation of a CT scan using a neural radiation field directly, and it is impossible to effectively distinguish surface shape, volume depth, and internal structure from 2D images.
Motivated by the well-known trade-off between distortion and perception in GAN methods, our method solves this problem by adding $\mathcal{L}_{\mathrm{r}}$ and $\mathcal{L}_{\mathrm{MSE}}$.
\begin{equation}
\mathcal{L}_{\mathrm{r}} = \mathbb{E}_{\boldsymbol{f} \sim D(\boldsymbol{p}), \boldsymbol{p} \sim \boldsymbol{P}} \left[ \frac{1}{w h d} \left\| \phi_i(\mathcal{G}(\boldsymbol{f})) - \phi_i(\mathcal{T}(\boldsymbol{p})) \right\|_2 \right]
\label{lr}
\end{equation}
 Where $\phi_i(\cdot)$ represents the output of the ith layer of the pre-trained VGG16 network model, ${w}$, ${h}$ and ${d}$ is the width, height, and depth of the feature space, $\mathcal{G}$ represents the processing of $\mathcal{D}_{\phi}$ intermediate feature $f$, and $\mathcal{T}$ is the processing of the truth image block.
\begin{equation}
\mathcal{L}_{\mathrm{MSE}} = \frac{1}{N} \sum_{i=1}^{N} (y_i - \hat{y}_i)^2
\label{lmse}
\end{equation}
Where $\hat{y}$ is the predicted value, $y$ is the true value, and $N$ is the number of predicted and true values.

\begin{equation}
\mathcal{L}_{\mathrm{gen}} = \lambda_1 \mathcal{L}_{r}(VGG16) + \lambda_2 \mathcal{L}_{MSE}(G)
\label{gen}
\end{equation}

In most cases, hyperparameters $ \lambda_1 $ and $ \lambda_2 $ are traditionally configured manually and heavily depend on human expertise, which can pose challenges in achieving the optimal trade-off between image blurriness and accuracy. 










\begin{figure*}[t] 
\centering
\includegraphics[width=0.9\textwidth]{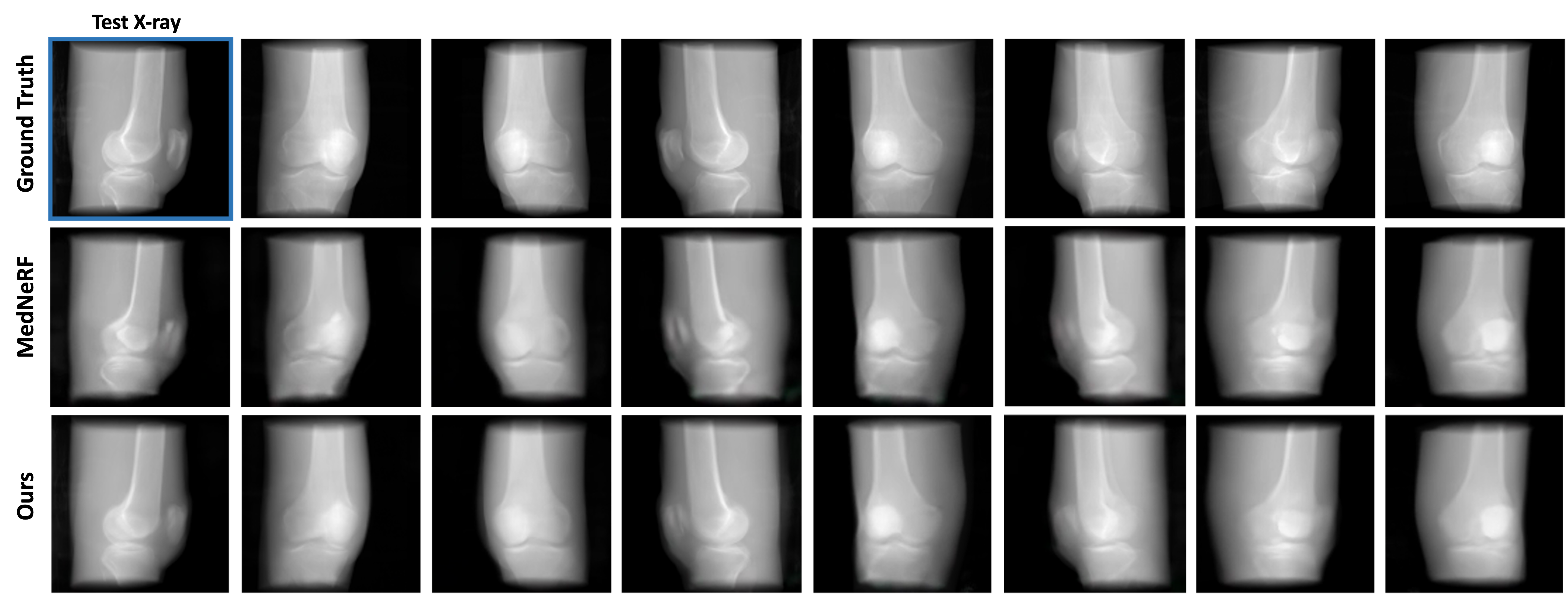}
\vspace{-2mm}
\caption{The single knee X-ray given yields a complete CT projection by our uncertainty-aware fine-tuned generator.}
\label{fig:result}
\vspace{-2mm}
\end{figure*}

\noindent\textbf{Uncertainty-aware Multitask Learning.}  In this section, we proposed a novel approach to automatically learn $\lambda_1$ and $\lambda_2$ in Eq. (\ref{gen}) based on task-dependent uncertainty \cite{kendall2018multi, zhao2020uncertainty}. 
The learning objective can be represented as 
\begin{equation}
    \begin{aligned}
        \mathcal{L}=\frac{1}{2\sigma_1^2}\mathcal{L}_{r}(VGG16)+\frac{1}{2\sigma_2^2}\mathcal{L}_{MSE}(G)+\log \sigma_1\sigma_2,
    \end{aligned}
\label{eq:loss}
\end{equation}
where $\sigma_1$ and $\sigma_2$ are the VGG16 and $G$ modules' observation noise parameters, which can capture how much noise they have in the outputs of each module. Minimizing $\mathcal{L}$ with respect to $\sigma_1$ and $\sigma_2$ can learn the relative weight of the losses adaptively \cite{hu2020learning}. The last term in Eq. (\ref{eq:loss}) is a regularizer, which can constrain the noise to be increased too much.

\noindent\textbf{Fine-tuning Generator.} In this phase, we reconstruct a complete CT projection across a full vertical rotation using a single X-ray image. 
In addition to adjusting the shape and appearance of latent vectors $z_s$ and $z_a$, we also fine-tuned the generator parameters using the test image. 
Furthermore, we also normalize the output to enhance the quality of the CT projections using $y' = \frac{y - \min(y)}{\max(y) - \min(y)} \times 255$.

\noindent\textbf{Testing.} In the single-view rendering stage,  we challenged our model by employing a single X-ray (referred to as a single view) for CT projection reconstruction over a full 360-degree range, each view taken at a 5-degree interval (See Fig. \ref{figov} (Right)). This testing phase is instrumental in gauging the model's performance across the entire angular spectrum.

\begin{table}[t]
\centering
\scalebox{0.8}{
\begin{tabular}{c|ll|cl}
\hline
\multirow{2}{*}{Method}           & \multicolumn{2}{c|}{Chest dataset}        & \multicolumn{2}{c}{Knee dataset}             \\ \cline{2-5} 
                                  & \multicolumn{1}{l|}{↑ PSNR (dB)} & ↑ SSIM & \multicolumn{1}{l|}{↑ PSNR (dB)}    & ↑ SSIM \\ \hline
\multicolumn{1}{l|}{MedNeRF \cite{corona2022mednerf}}      & \multicolumn{1}{c|}{28.37}            &      0.462     & \multicolumn{1}{c|}{30.93}          &     0.670   \\
 \multicolumn{1}{c|}{\textbf{Ours}} & \multicolumn{1}{c|}{\textbf{28.41}}            &    \textbf{0.484}    & \multicolumn{1}{c|}{\textbf{31.62}} &     \textbf{0.723}  \\ \hline
\end{tabular}
}
\vspace{-2mm}
\caption{Quantitative results based on PSNR and SSIM of rendered X-ray projections with single-view X-ray input.}
\label{tab:psnr}
\end{table}

\section{Experiments}
\label{sec:copyright}


\noindent\textbf{Dataset.} To ensure comparability with existing methodologies, we leveraged the identical CT dataset employed in the MedNeRF paper to produce DRRs. 
The dataset for generating DRRs comprised 20 CT chest scans as referenced in \cite{tsai2021medical} and \cite{clark2013cancer}, in addition to 5 CT knee scans documented in \cite{harris2016combined} and \cite{ali2016validation}. 
These scans represented a diverse array of patients, encompassing various contrast modalities and encapsulating both typical and atypical anatomical structures. 
The DRR generation process involved the emulation of the rotation of a radiation source and imaging panel around the vertical axis, resulting in the creation of DRRs with a resolution of 128 $\times$ 128 at five-degree intervals. 
Consequently, each object produced a total of 72 distinct DRRs. 


\begin{table}[t]
\centering
\scalebox{0.65}{
\begin{tabular}{l|ll|ll}
\hline
\multicolumn{1}{c|}{\multirow{2}{*}{Method}} & \multicolumn{2}{c|}{Chest dataset} & \multicolumn{2}{c}{Knee dataset} \\ \cline{2-5} 
\multicolumn{1}{c|}{} & \multicolumn{1}{l|}{$\downarrow$ FID ($\mu \pm \sigma$)} & $\downarrow$ KID ($\mu \pm \sigma$) & \multicolumn{1}{l|}{$\downarrow$ FID ($\mu \pm \sigma$)} & $\downarrow$ KID ($\mu \pm \sigma$) \\ \hline
GRAF \cite{schwarz2020graf} & \multicolumn{1}{l|}{68.25 $\pm$ 0.954} & 0.053 $\pm$ 0.0008 & \multicolumn{1}{c|}{76.70 $\pm$ 0.302} & 0.058 $\pm$ 0.0001 \\
\multicolumn{1}{r|}{pixelNeRF \cite{yu2021pixelnerf}} & \multicolumn{1}{l|}{112.96 $\pm$ 2.356} & 0.084 $\pm$ 0.0014 & \multicolumn{1}{c|}{166.40 $\pm$ 2.245} & 0.158 $\pm$ 0.0010 \\
GIRAFFE \cite{niemeyer2021giraffe} & \multicolumn{1}{l|}{66.42 $\pm$ 0.706} & 0.064 $\pm$ 0.0012 & \multicolumn{1}{l|}{184.34 + 0.281} & 0.064 $\pm$ 0.0012 \\
MedNeRF \cite{corona2022mednerf} & \multicolumn{1}{l|}{62.40 $\pm$ 0.352} & 0.049 $\pm$ 0.0008 & \multicolumn{1}{l|}{76.41 $\pm$ 2.153} & 0.057 $\pm$ 0.0006 \\
\textbf{Ours} & \multicolumn{1}{l|}{\textbf{60.25 $\pm$ 0.642}} & \textbf{0.043 $\pm$ 0.0011} & \multicolumn{1}{l|}{\textbf{70.73 $\pm$ 1.665}} & \textbf{0.041 $\pm$ 0.0012} \\ \hline
\end{tabular}
}
\vspace{-3mm}
\caption{Comparing with other methods.}
\label{tab:fid}
\end{table}

\vspace{-3mm}

\noindent\textbf{Evaluation Metrics.} We evaluate our model's performance in CT projection reconstruction using two key image quality metrics: Peak Signal-to-Noise Ratio (PSNR) and Structural Similarity Index (SSIM). PSNR emphasizes luminance accuracy, measuring differences between original and model-generated images. SSIM provides a comprehensive assessment, considering luminance, contrast, and structural similarity. These metrics are crucial for accurately assessing our model's performance in CT projection reconstruction, ensuring reliability in our research findings.

\noindent\textbf{Comparison with baseline models.}
We first compared it with the baseline MedNeRF \cite{corona2022mednerf}. Table \ref{tab:psnr} displays our results, which were assessed using PSNR and SSIM. 
As shown in Fig. \ref{fig:result}, our model can better find the balance between blurriness and accuracy in the quality of generated image results.
Our findings highlight the effectiveness of Automatic Weighted Loss in improving the clarity of reconstructed images and revealing finer details within bone structures. Moreover, our model demonstrates the ability to consistently generate CT projections from multiple consecutive viewpoints.

We also evaluate our model on the task of 2D rendering and compare it to \cite{schwarz2020graf}, \cite{yu2021pixelnerf}, and \cite{niemeyer2021giraffe} baseline, Table \ref{tab:fid} presents a comparison of image quality using the Frechet Inception Distance (FID) and Kernel Inception Distance (KID) metrics, where lower values indicate better quality. When optimizing pixelNeRF and GIRAFFE on our datasets, it yields notably inferior results that do not match the performance of the GRAF baseline or our model. In contrast, our model surpasses the baseline models in terms of FID and KID metrics across all datasets.

\begin{table}[t]
\centering
\scalebox{0.8}{
\begin{tabular}{cc|cc}
\hline
\multicolumn{2}{c|}{Task Weights}                                                                              & \multirow{2}{*}{PSNR} & \multirow{2}{*}{SSIM} \\
$\mathcal{L}_{r}$                                                    & $\mathcal{L}_{MSE}$                                                     &                       &                       \\ \hline
0.6                                                    & 0.4                                                   & 30.93                 & 0.67                  \\
0.5                                                    & 0.5                                                   & 30.67                & 0.61                   \\
0.4                                                    & 0.6                                                   & 31.26                 & 0.70                      \\ \hline
\multicolumn{2}{c|}{\textbf{\begin{tabular}[c]{@{}c@{}}Learned weights \\ with task uncertainty\end{tabular}}} & \textbf{31.62}        & \textbf{0.72}         \\ \hline
\end{tabular}
}
\vspace{-2mm}
\caption{ Comparison of Loss Weights on Knee Dataset.}
\label{tab:Ablation}
\vspace{-5mm}
\end{table}

\noindent\textbf{Ablation Study.} Due to the balance between image clarity and accuracy being a crucial concern in GAN network structures, we introduce Mean Squared Error (MSE) loss during the generation phase in an attempt to address this issue \cite{hu2023rank}. The model's performance is highly sensitive to the choice of weight parameters. A common approach for combining multiple objective losses is to simply assign weights to each objective and sum them up. However, this method heavily relies on manual expertise and can be quite costly in terms of experimentation, often taking many days to fine-tune.
In our UMedNeRF, we utilize Autoweightloss, which is capable of automatically learning the optimal weights for the task. As shown in Table \ref{tab:Ablation}, we employ various sets of loss weight parameters for ablation studies to assess the effectiveness of our approach. Experimental results demonstrate that UMedNeRF efficiently addresses the balance issue in generating images.

\section{Conclusion}

Our research proposes a novel deep learning architecture based on neural radiance fields for learning continuous representations of CT scans from 2D input X-rays. 
Our proposed UMedNeRF solves the distortion and perception tradeoff common in GAN methods to generate images through Autoweight loss and uses a single X-ray to achieve a complete CT projection reconstruction. 
The experimental results show that UMedNeRF has a significant improvement in the quality of CT projections compared with other neural radiance field methods. 
Although this model cannot completely replace CT at the current stage, this method can greatly reduce the radiation dose received by patients in clinical examinations and reduce the time and economic costs, which has great research potential in bone trauma and surgery.
\label{sec:Conclusion}


\section{COMPLIANCE WITH ETHICAL STANDARDS} 
This research study was conducted retrospectively using human subject data made available in open access by (Source information). Ethical approval was not required as confirmed by the license attached with the open access data.

\section{Acknowledgments}

This work was supported in part by the National Natural Science Foundation of China under Grants 42375148, Sichuan province Key Technology Research and Development project under Grant 2023YFG0305.

Xin Wang is supported by University at Albany Start-up Grant.
\label{sec:acknowledgments}

\small
\bibliographystyle{IEEEbib}
\bibliography{refs}

\begin{thebibliography}{10}

\bibitem{lo2012extraction}
Pechin Lo, Bram Van~Ginneken, Joseph~M Reinhardt, et~al.,
\newblock ``Extraction of airways from ct (exact'09),''
\newblock {\em IEEE Transactions on Medical Imaging}, vol. 31, no. 11, 2012.

\bibitem{wang2024artificial}
Xin Wang and Hongtu Zhu,
\newblock ``Artificial intelligence in image-based cardiovascular disease
  analysis: A comprehensive survey and future outlook,''
\newblock {\em arXiv:2402.03394}, 2024.

\bibitem{maken20232d}
Payal Maken and Abhishek Gupta,
\newblock ``2d-to-3d: A review for computational 3d image reconstruction from
  x-ray images,''
\newblock {\em Archives of Computational Methods in Engineering}, 2023.

\bibitem{wang2023deep}
Xin Wang, Ziwei Luo, Jing Hu, Chengming Feng, Shu Hu, et~al.,
\newblock ``Deep reinforcement learning for image-to-image translation,''
\newblock {\em arXiv:2309.13672}, 2023.

\bibitem{mildenhall2021nerf}
Ben Mildenhall, Pratul~P Srinivasan, et~al.,
\newblock ``Nerf: Representing scenes as neural radiance fields for view
  synthesis,''
\newblock 2021, vol.~65, ACM New York, NY, USA.

\bibitem{wang2024neural}
Xin Wang, Shu Hu, Heng Fan, Hongtu Zhu, and Xin Li,
\newblock ``Neural radiance fields in medical imaging: Challenges and next
  steps,'' 2024.

\bibitem{corona2022mednerf}
Abril Corona-Figueroa, Jonathan Frawley, et~al.,
\newblock ``Mednerf: Medical neural radiance fields for reconstructing 3d-aware
  ct-projections from a single x-ray,''
\newblock in {\em EMBC}. IEEE, 2022.

\bibitem{kendall2018multi}
Alex Kendall, Yarin Gal, and Roberto Cipolla,
\newblock ``Multi-task learning using uncertainty to weigh losses for scene
  geometry and semantics,''
\newblock in {\em CVPR}, 2018.

\bibitem{zhao2020uncertainty}
Xujiang Zhao, Feng Chen, Shu Hu, and Jin-Hee Cho,
\newblock ``Uncertainty aware semi-supervised learning on graph data,''
\newblock {\em NeurIPS}, vol. 33, pp. 12827--12836, 2020.

\bibitem{hu2020learning}
Shu Hu, Yiming Ying, Xin Wang, and Siwei Lyu,
\newblock ``Learning by minimizing the sum of ranked range,''
\newblock {\em NeurIPS}, vol. 33, pp. 21013--21023, 2020.

\bibitem{tsai2021medical}
E.~B. Tsai et~al.,
\newblock ``Medical imaging data resource center (midrc) - rsna international
  covid open research database (ricord) release 1b - chest ct covid,''
\newblock 2021.

\bibitem{clark2013cancer}
K.~Clark, B.~Vendt, K.~Smith, J.~Freymann, et~al.,
\newblock ``The cancer imaging archive (tcia): Maintaining and operating a
  public information repository,''
\newblock 2013, vol.~26.

\bibitem{harris2016combined}
Michael~D. Harris, Adam~J. Cyr, Azhar~A. Ali, et~al.,
\newblock ``A combined experimental and computational approach to
  subject-specific analysis of knee joint laxity,''
\newblock 2016, vol. 138.

\bibitem{ali2016validation}
Azhar~A. Ali, Sami~S. Shalhoub, Adam~J. Cyr, et~al.,
\newblock ``Validation of predicted patellofemoral mechanics in a finite
  element model of the healthy and cruciate-deficient knee,''
\newblock 2016.

\bibitem{schwarz2020graf}
Katja Schwarz, Yiyi Liao, Michael Niemeyer, and Andreas Geiger,
\newblock ``Graf: Generative radiance fields for 3d-aware image synthesis,''
\newblock 2020, vol.~33.

\bibitem{yu2021pixelnerf}
Alex Yu, Vickie Ye, et~al.,
\newblock ``pixelnerf: Neural radiance fields from one or few images,''
\newblock in {\em CVPR}, 2021.

\bibitem{niemeyer2021giraffe}
Michael Niemeyer and Andreas Geiger,
\newblock ``Giraffe: Representing scenes as compositional generative neural
  feature fields,''
\newblock in {\em CVPR}, 2021.

\bibitem{hu2023rank}
Shu Hu, Xin Wang, and Siwei Lyu,
\newblock ``Rank-based decomposable losses in machine learning: A survey,''
\newblock {\em IEEE Transactions on Pattern Analysis and Machine Intelligence},
  2023.

\end{thebibliography}

\end{document}